\documentclass[runningheads,a4paper]{llncs}

\usepackage{amssymb}
\usepackage{graphicx}

\usepackage{url}
\urldef{\mailsa}\path|amironov66@gmail.com|
\urldef{\mailsb}\path|anna.kramer, leonie.kunz, christine.reiss, nicole.sator,|
\urldef{\mailsc}\path|erika.siebert-cole, peter.strasser, lncs}@springer.com|    
\newcommand{\keywords}[1]{\par\addvspace\baselineskip
\noindent\keywordname\enspace\ignorespaces#1}

\begin{document}

\def\dey#1#2{#1 (#2)}
\def\deyc#1#2{#1 \cdot  #2}
\def\bcy{\begin{array}{ccc}}

\def\ral#1{\;\mathop{\longrightarrow}\limits^{#1}\;}
\def\bc{\begin{center}\begin{tabular}{l}}
\def\ec{\end{tabular}\end{center}}
\def\modn#1{\mathop{=}\limits_{#1}}
\def\modnop#1{\mathop{#1}\limits_{n}}
\def\rar{\mathop{\in}\limits_{r}}
\def\und#1{\mathop{=}\limits_{#1}}
\def\skobq#1{\langle\!| #1 |\!\rangle}
\def\redeq{\;\mathop{\approx}\limits^{r}\;}
\def\reduc{\;\mathop{\mapsto}\limits^{r}\;}
\def\pt{\;\mathop{+}\limits_{\tau}\;}
\def\sost{\begin{picture}(0,0)\put(0,3){\circle*{4}}
\end{picture}}
\def\sosto{\begin{picture}(0,0)\put(0,0){\circle*{4}}
\end{picture}}
\def\bi{\begin{itemize}}
\def\pa{\,|\,}
\def\oc{\;\mathop{\approx}\limits^{+}\;}
\def\p#1#2{(\;#1\;,\;#2\;)}
\def\mor#1#2#3{\by #1&\pright{#2}&#3\ey}
\def\ei{\end{itemize}}
\def\bn{\begin{enumerate}}
\def\en{\end{enumerate}}
\def\i{\item}
\def\a{\forall\;}

\def\l#1{\langle #1 \rangle}

\def\bigset#1#2{\left\{\by #1 \left| \by #2 \ey\right\}\ey\right.}
\def\p{\leftarrow}

\def\plongright#1{
  \begin{picture}(40,8)
  \put (-5,3){\vector(1,0){50}}
  \put (20,8){\makebox(1,1){$\scriptstyle #1$}}
  \end{picture} }

\def\plongleft#1{
  \begin{picture}(40,8)
  \put (45,3){\vector(-1,0){50}}
  \put (20,8){\makebox(1,1){$\scriptstyle #1$}}
  \end{picture} }

\def\pse#1#2{
  \begin{picture}(40,8)
  \put (-5,-5){\vector(2,-1){50}}
  \put (45,-5){\vector(-2,-1){50}}
  \put (-5,-12){\makebox(1,1)[r]{$\scriptstyle #1$}}
  \put (45,-12){\makebox(1,1)[l]{$\scriptstyle #2$}}
  \end{picture} }

\def\und#1{\mathop{=}\limits_{#1}}
\def\redeq{\;\mathop{\approx}\limits^{r}\;}
\def\reduc{\;\mathop{\mapsto}\limits^{r}\;}
\def\oc{\mathop{\approx}\limits^{+}}
\def\sost{\begin{picture}(0,0)\put(0,0){\circle*{4}}
\end{picture}}
\def\bi{\begin{itemize}}
\def\pa{\,|\,}
\def\oo{\;\mathop{\approx}\limits^{c}\;}
\def\p#1#2{(\;#1\;,\;#2\;)}
\def\mor#1#2#3{\by #1&\pright{#2}&#3\ey}
\def\ei{\end{itemize}}
\def\bn{\begin{enumerate}}
\def\en{\end{enumerate}}
\def\i{\item}
\def\bigset#1#2{\left\{\by #1 \left| \by #2 \ey\right\}\ey\right.}
\def\p{\leftarrow}
\def\buffer{{\it Buffer}}
\def\eam{\mathbin{{\mathop{=}\limits^{\mbox{\scriptsize def}}}}}
\def\be#1{\begin{equation}\label{#1}}
\def\ee{\end{equation}}
\def\re#1{(\ref{#1})}

\def\bn{\begin{enumerate}}
\def\en{\end{enumerate}}
\def\bi{\begin{itemize}}
\def\ei{\end{itemize}}
\def\i{\item}
\def\c#1{\left\{\begin{array}{lllll}#1\end{array}\right\}}
\def\d#1{\left[\begin{array}{lllll}#1\end{array}\right]}
\def\b#1{\left(\begin{array}{lllll}#1\end{array}\right)}
\def\ra#1{\;\mathop{\to}\limits^{#1}\;}
\def\leqd{\;\mathop{<}\limits_{2}\;}
\def\diagrw#1{{
  \def\normalbaselines{\baselineskip20pt \lineskip3pt \lineskiplimit3pt }
  \matrix{#1}}}

\def\blackbox{\vrule height 7pt width 7pt depth 0pt}
\def\pu#1#2{
\mbox{$\!\!\begin{picture}(0,0)
\put (-#1,-#2){\line(1,0){#1}}
\put (-#1,-#2){\line(0,1){#2}}
\put (#1,#2){\line(-1,0){#1}}
\put (#1,#2){\line(0,-1){#2}}
\put (-#1,#2){\line(1,0){#1}}
\put (-#1,#2){\line(0,-1){#2}}
\put (#1,-#2){\line(-1,0){#1}}
\put (#1,-#2){\line(0,1){#2}}
\end{picture}$}
}

\def\pright#1{
  \begin{picture}(20,8)
  \put (-5,3){\vector(1,0){30}}
  \put (9,8){\makebox(1,1){$\scriptstyle #1$}}
  \end{picture} }

\def\by{\begin{array}{llllllllllllll}}
\def\ey{\end{array}}

\mainmatter  

\title{A New Method of Verification \\of Functional Programs}



\author{Andrew M. Mironov%
%
}

\institute{Moscow State University
\\$\;$\\
\mailsa\\
}

\maketitle

\begin{abstract}
In the paper the problem of verification of functional programs (FPs) over strings is considered,
where specifications of properties 
of FPs are defined by other FPs,
and a FP $\Sigma_1$ meets a specification defined by another FP $\Sigma_2$ iff a composition of functions defined by the FPs
$\Sigma_1$ and $\Sigma_2$ is equal 
to the constant 1.
We introduce a concept of a state diagram of  a  FP,
and reduce the verification problem to the problem
of an analysis  of the state diagrams of FPs. The proposed approach is illustrated by the example of 
verification of a sorting program.
\keywords{functional program,
state diagram,
verification}
\end{abstract}


\section{Introduction}

The problem of program verification is one 
of the main problems 
of theoretical computer science. 
For various classes of programs 
there are used various verification methods.
For example, for a verification of sequential
programs there are used  Floyd's
inductive assertions method 
\cite{floyd}, Hoare logic \cite{hoare}, etc.
For verification of parallel
and distributed programs
there are used methods based on
a calculus of communicating systems
(CCS) and $\pi$-calculus \cite{milner1},
\cite{milner2}, a theory of communicating
sequential processes
(CSP) and its generalizations \cite{csp},
\cite{sep}, temporal logic and
model checking \cite{peled}, process algebra
\cite{pa}, Petri nets \cite{petri}, etc.

Methods of verification  of functional programs
(FPs) are developed not so completely,
as verification methods for sequential 
and parallel programs.
Main methods of verification of FPs
are computational induction and 
structural induction \cite{manna}.
Disadvantages of these methods are 
related to difficulties to construct 
formal proofs of program correctness.
Among other methods of verification of FPs
it should be noted a method based on
reasoning with datatypes and abstract
interpretation
through type inference
\cite{rybal},
a model checking method to verify FPs
\cite{8},
\cite{14},
methods based on flow analysis
\cite{5}
methods based on the concept of a multiparametric tree transducer 
\cite{9}.

In this article we consider FPs as
systems of algebraic equations over strings.
We introduce a concept of 
a state diagram for such FPs and
 present the verification method
based on the state diagrams.
The main advantages of our approach
in comparison with 
all the above approaches to verification of FPs
are related to
 the fact that our approach allows
 to present proofs of correctness
of FPs in the form of simple properties 
of their state diagrams.

The basic idea of our approach is
the following.
We assume that a specification of properties
of a FP under verification
$\Sigma_1$
is defined by another FP $\Sigma_2$,
whose input is equal to 
the output of $\Sigma_1$,
i.e. we consider FP
$\Sigma_1 \circ \Sigma_2$, which is
a composition
$\Sigma_1$ and $\Sigma_2$.
We say that a FP
$\Sigma_1$ is correct with respect to 
the specification $\Sigma_2$ iff
the input-output map 
$f_{\Sigma_1 \circ \Sigma_2}$,
which corresponds to the FP
$\Sigma_1 \circ \Sigma_2$
(i.e. $f_{\Sigma_1 \circ \Sigma_2}$ is
a composition of the input-output  maps
corresponded to
$\Sigma_1$ and $\Sigma_2$)
has an output value 1 on all its input values.
We reduce the
problem of a proving
the statement
$f_{\Sigma_1 \circ \Sigma_2} = 1$
to  the problem of an 
analysis of a state diagram for the FP
$\Sigma_1 \circ \Sigma_2$.

The proposed method of verification of FPs
is illustrated 
by an example of verification of a sorting FP. 
At first, we present a complete proof
of correctness of this FP
by structural induction. This is done
for a comparison of a complexity of 
a manual verification of the FP 
on the base of 
the structural induction method, and 
a complexity of 
the proposed method of automatic 
verification of FPs.
At second, we present a correctness proof 
 of the FP by the method based on 
 constructing its state diagram.
The proof by the second method is
significantly shorter, 
and moreover, it can be generated
automatically.
This demonstrates the benefits of
the proposed method of 
verification of FPs
in comparison with the manual 
verification based on the
structural induction method.

\section{Main concepts} 

\subsection{Terms} 

We assume that there are given sets
\bi
\i ${\cal D}$ of {\bf values}, 
which is the union
${\cal D}_{\bf C}\cup {\cal D}_{\bf S}$,
where
\bi
\i  elements of ${\cal D}_{\bf C}$ are called
{\bf  symbols},
and
\i  elements of ${\cal D}_{\bf S}$ are called
{\bf symbolic strings} (or briefly
{\bf strings}), and
each string from ${\cal D}_{\bf S}$
is a finite (maybe empty) sequence
of elements of ${\cal D}_{\bf C}$, \ei
\i
${\cal X}$ of {\bf data variables} 
(or briefly {\bf variables})
\i ${\cal C}$ of {\bf constants},
\i ${\cal F}$ of {\bf functional
symbols (FSs)}, and
\i $\Phi$ of {\bf functional
variables}
\ei
where each element $m$ of any of the
above sets is associated with a {\bf type}
designated by the notation $type(m)$, and
\bi
\i if $m\in{\cal D}\cup
{\cal X}\cup{\cal C}$,
then $type (m)\in \{{\bf C}, {\bf S}\}$,
\i if $m\in{\cal F}\cup \Phi$, then 
$type(m)$ is a notation of the form
$t_1\times \ldots\times t_n\to t$, where
$t_1,\ldots, t_n, t \in \{{\bf C}, {\bf S}\}$.
\ei

If $d\in {\cal D}_{\bf C}$, 
then $type (d) = {\bf C}$,
and if $d\in {\cal D}_{\bf S}$, 
then  $type (d) = {\bf S}$.

Each constant $c\in {\cal C}$
corresponds to an element of 
${\cal D}_{type(c)}$, called
a {\bf value} of this constant.
The notation $\varepsilon$ 
denotes a constant of the 
type {\bf S}, whose value
is an empty string.
We  assume that
$\varepsilon $ is the only
constant of the type {\bf S}.

Each FS $f\in {\cal F}$ corresponds to a 
partial function of the form
${\cal D}_{t_1}\times\ldots \times
{\cal D}_{t_n} \to {\cal D}_t$, where
  $$type(f)=
t_1\times \ldots\times t_n\to t.$$
This function is
denoted by the same symbol $f$.

Below we list some of the FSs 
which belong to 
${\cal F}$.
Beside each FS we point out 
(with a colon) its type.

\bn
\i $head: {\bf S} \to {\bf C}$.
   The function $head$ is defined for
   non-empty string,
   it maps each non-empty string
   to its first element.
\i $tail: {\bf S} \to {\bf S}$.
   The function $tail$ is defined for
   non-empty string,
   it maps each non-empty string $u$ 
   to a string (called a {\bf tail} of $u$)
   derived from $u$ by removal 
   of its first element.
\i $conc: {\bf C}\times {\bf S}\to {\bf S}$.
   For each pair
   $(a,u)\in {\cal D}_{\bf C}\times 
   {\cal D}_{\bf S}$ 
   the string $conc(a,u)$ is
   obtained from $u$ by adding 
   the symbol $a$ before.
\i $empty: {\bf S}\to {\bf C}$.
   Function $empty$ maps 
   empty string  to the symbol 1, 
   and
   each non-empty string to the symbol 0.

\i $= :{\bf C}\times {\bf C}\to {\bf C}$.
The value of the function $=$ on 
the pair $(u, v)$ is equal to
1 if $u = v$,
and 0 otherwise.

\i $\leq :{\bf C}\times {\bf C}\to {\bf C}$.
We assume that ${\cal D}_{\bf C}$
is linearly ordered set, and
the value of the function $\leq$ 
on the pair $(u, v)$ is equal to
1 if $u\leq v$, and 0 otherwise.

\i Boolean FSs:
$\neg: {\bf C}\to {\bf C}$,
$\wedge: {\bf C}\times {\bf C}\to {\bf C}$,
etc., corresponding functions 
are standard boolean functions
on the arguments 
0 and 1
(i.e. $\neg (1) = 0, $, etc.)
and are not defined on other arguments.

\i ${\it if\_then\_else}\;: {\bf C}\times t\times
    t\to t$, where $t={\bf C}$ or ${\bf S}$
(i.e. the notation ${\it if\_then\_else}$ 
denotes two FSs), and
    functions corresponding to both
    FSs are defined by the same way:
   $${\it if\_then\_else}\;(a,u,v)\eam
   \left\{\by u,&\mbox{if $a=1$}\\
   v,&\mbox{otherwise}. \ey\right.$$
\en

A concept of a {\bf term} is defined inductively.
Åach term $e$ is associated with 
a certain type
 $type(e) \in \{{\bf C}, {\bf S}\}$.
Each data variable and each constant 
is a term, a type of which is the same
as the type of this variable or constant.
If
$e_1,\ldots, e_n$ is a list of terms and
$g$ is a FS or a functional
variable such that $$type(g) = \;\;
type(e_1)\times \ldots\times type(e_n)\to t$$
then the 
notation $g(e_1,\ldots, e_n)$
is a term of the type $t$.

We shall notate
terms $$\by
head(e),\; tail(e),\;conc(e_1, e_2),\;
   empty(e),\\
    =(e_1,e_2),\;
\leq (e_1,e_2),\;
       {\it if\_then\_else}\;(e_1,e_2,e_3)\ey$$
       in the form
\bc
   $e_h$, $e_t$, $e_1 e_2$, 
   $e = \varepsilon$,
         $e_1 = e_2$,
         $e_1 \leq e_2$,
      $e_1\;?\;e_2:e_3$\ec
      respectively.
      Terms containing boolean FSs
will be notated as 
in mathematical texts (i.e. in the form
      $e_1\wedge e_2$, etc.).
Terms of the form
$e_1\wedge \ldots\wedge
e_n$ can also be notated as
$\{e_1, \ldots,  e_n\}$.

\subsection{A concept of a functional program over strings}

A {\bf functional program over strings}
(referred below  
as a {\bf functional program (FP)})
is a set  $\Sigma$ of functional
equations of the form 
\be{dfgdsfgdsr5rtt6}
\left\{\by\varphi_1(x_{11},\ldots, x_{1n_1})=e_1\\
\ldots\\
\varphi_m(x_{m1},\ldots, x_{mn_m})=e_m
\ey\right.\ee
where
$\varphi_1, \ldots, \varphi_m$ are
distinct functional variables, and
for each 
$i=1,\ldots, m\;\;
\varphi_i(x_{i1},\ldots,x_{in_i})$
      and $e_i$ are terms of 
      the same type,
such that
  $$X_{e_i}=\{x_{i1},\ldots,x_{in_i}\},\quad
  \Phi_{e_i}\subseteq\{\varphi_1,\ldots,
\varphi_m\}.$$

We shall use
the notation $\Phi_\Sigma$
for 
the set of all functional variables
occurred in $\Sigma$.

FP \re{dfgdsfgdsr5rtt6}
specifies a list 
\be{dfasdfasd}(f_{\varphi_1},\ldots, f_{\varphi_m})\ee
of functions corresponded to the 
functional variables from $\Phi_\Sigma$,
which is the least 
(in the sense of an order on lists of partial 
functions, described in \cite{manna})
solution of \re{dfgdsfgdsr5rtt6}
(this list is called
a {\bf least fixed
point (LFP)} of the FP 
\re{dfgdsfgdsr5rtt6}).
Values of these functions can be
calculated by a standard recursion.
We assume that for each
FP under consideration
all components of its LFP 
are total functions.
First function in the list \re{dfasdfasd}
(i.e. $f_{\varphi_1}$) is denoted by 
$f_\Sigma$,  and is called
a {\bf function corresponded to $\Sigma$}.
If $\Sigma$ has the form
\re{dfgdsfgdsr5rtt6}, then
$type(\Sigma)$ 
denotes the type $type(e_1)$.

\section{Example of specification 
and verification of a FP}

\subsection{Example of a FP}
\label{4.3.2}

Consider the following FP:
\be{fdsgdsgdsr}
\by
\by {\bf sort}(x) = (x = \varepsilon) ? \;\varepsilon
 :  {\bf insert}(x_h, {\bf sort}(x_t))\ey\\
\by  {\bf insert}(a, y) = (y = \varepsilon)& ?\;a  \varepsilon\\
&: (a \leq  y_h) &?\; a  y\\
&&: y_h   \;{\bf insert}(a, y_t)\ey\ey\ee
This FP defines a function
of string sorting.
The FP consists of two equations, 
which define the following functions:
\bi
\i ${\bf sort}: {\bf S} \to {\bf S}$ is a main
  function, and
\i ${\bf insert}: {\bf C}\times {\bf S}\to {\bf S}$ is an auxiliary function, which
maps a pair $(a,y)\in {\bf C}\times
 {\bf S}$ to the string derived by
 an insertion of the symbol $a$
to the string $y$, with the following 
property: if the string 
$y$ is ordered, then
the string ${\bf insert}(a,y)$ also
is ordered. \\
(we say that a string is ordered,
if its components form a nondecreasing
sequence).
\ei

\subsection{Example of a specification
of a FP} \label{sdfsadfsad}

One of correctness properties of 
 FP \re{fdsgdsgdsr}
it the following:
$\forall\,x\in {\bf S}$ 
the string ${\bf sort}(x)$ is ordered.
This property
can be  described formally as follows.
Consider a FP 
defining
a function {\bf ord}  of  string
ordering checking:
\be{sdfgfdsgsdfgdsfgsrrr}
\by
{\bf ord}(x) = \\
=(x = \varepsilon) &?\; 1\\
&: (x_t = \varepsilon) &? \;1\\
&&: (x_h \leq (x_t)_h) &?\; {\bf ord}(x_t)\\
&&&: 0\ey\ee

The function {\bf ord} allows
 describe
the above property of correctness
as the following
mathematical statement:
\be{4.21}\forall\,  x \in {\bf S}\quad
{\bf ord}({\bf sort}(x)) = 1 \ee

\subsection{Example of a verification of a FP}

The problem of verification 
of the correctness property 
of FP \re{fdsgdsgdsr}
consists of a formal proof of \re{4.21}.
This proposition can be proved
like an ordinary mathematical
theorem, for example using the method
of mathematical induction.
For example, a proof 
of this proposition can be the following.

If $x = \varepsilon$, then,
according to  first equation of  system
\re{fdsgdsgdsr}, the equality
 ${\bf sort}(x) = \varepsilon$ holds, 
 and therefore
$${\bf ord}({\bf sort}(x)) = {\bf ord}(\varepsilon) = 1.$$

Let $x \neq  \varepsilon$.
We prove \re{4.21} for 
this case by
induction. 
Assume that for each  string $y$,
which is shorter than $x$,
the equality 
$${\bf ord}({\bf sort}(y)) = 1$$
holds. Prove that this implies the equality
\be{dfasdfs}{\bf ord}({\bf sort}(x)) = 1.\ee

\re{dfasdfs} is equivalent to the equality
\be{4.22}
{\bf ord}( \;{\bf insert}(x_h, {\bf sort}(x_t))) = 1 \ee
By the induction hypothesis, the equality
$${\bf ord}({\bf sort}(x_t)) = 1$$
holds, and this
implies \re{4.22} on the reason 
of the following lemma. \\

{\bf Lemma}.

The following implication holds:
\be{4.23}
{\bf ord}(y) = 1 \quad\Rightarrow\quad
 {\bf ord}( {\bf insert}(a,  y)) = 1 \ee

{\bf Proof}.

We prove the lemma by induction 
on the length of $y$.

If $y = \varepsilon$, then the right side of\re{4.23} has the form
$${\bf ord}(a  \varepsilon) = 1$$
which is true by definition {\bf ord}.

Let $y \neq  \varepsilon$, and for each
string $z$, which is shorter than $y$,
the following implication holds:
\be{4.24}
{\bf ord}(z) = 1 \quad\Rightarrow\quad 
{\bf ord}( {\bf insert}(a, z)) = 1 \ee

Let $c \eam  y_h$, $d \eam y_t$.

\re{4.23} has the form
\be{4.25}
{\bf ord}(c  d) = 1 \quad
\Rightarrow\quad
 {\bf ord}( {\bf insert}(a,  c   d)) = 1\ee

To prove the implication \re{4.25}
it is necessary to prove that
if 
${\bf ord}(c  d) = 1$, then 
the following implications hold:
\bi
\i[(a)] $a \leq c \quad\Rightarrow\quad 
{\bf ord}(a   (c   d)) = 1$,
\i[(b)] $c < a \quad\Rightarrow\quad  {\bf ord}(c   
 \;{\bf insert}(a,  d)) = 1$.
\ei

(a) holds because $a \leq c$ implies
$${\bf ord}(a   (c   d)) = {\bf ord}(c   
d) = 1.$$

Let us prove (b).

\bi
\i $d = \varepsilon$. 
In this case,  right side of (b) has the form
\be{4.26}
{\bf ord}(c   (a   \varepsilon)) = 1\ee

\re{4.26} follows from $c <a$.
\i $d \neq  \varepsilon$. 
Let $p \eam  d_h$, 
$q \eam  d_t$.

In this case, it is necessary to prove that if $c < a$, then
\be{4.27}
{\bf ord}(c     \;{\bf insert}(a, p   q)) = 1 \ee

\bn
\i if $a \leq p$, then 
\re{4.27} has the form
\be{4.28}
{\bf ord}(c    (a   (p   q))) = 1\ee

Since $c < a\leq  p$, then \re{4.28}
follows from the equalities
$$\by
{\bf ord}(c   (a   (p   q))) = {\bf ord}(a   (p   q)) = {\bf ord}(p   q) =\\
= {\bf ord}(c   (p   q)) = {\bf ord}(c   d) = 1\ey$$
\i if $p <a$, then \re{4.27} has the form
\be{4.29}
{\bf ord}(c   (p    \;{\bf insert}(a, q))) = 1 \ee

Since, by assumption,
$${\bf ord}(c   d) = {\bf ord}(c   (p   q)) = 1$$
then $c \leq p$, 
and therefore \re{4.29} can be rewritten as
\be{4.30}
{\bf ord}(p    \;{\bf insert}(a, q)) = 1\ee 

If $p <a$, then
$$ {\bf insert}(a, d) =  {\bf insert}(a, p   q) 
= p    \;{\bf insert}(a, q)$$
therefore \re{4.30} can be rewritten as
\be{4.31}{\bf ord}( {\bf insert}(a, d)) = 1\ee 
\re{4.31} follows by the induction hypothesis for
the Lemma (i.e., 
from the implication \re{4.24},
where $z \eam d$) from
the equality
$${\bf ord} (d) = 1$$
which is justified
by the chain of equalities
$$\by 1 = {\bf ord}(c   d) = {\bf ord}(c   
(p    q)) = \quad(\mbox{since }c \leq p)\\
= {\bf ord}(p   q) = {\bf ord}(d).\quad\blackbox\ey
$$
\en
\ei

From the above example we can see that even
for the simplest FP, which consists of 
several lines,
a proof of its correctness is
not trivial mathematical reasoning,
it is difficult to check it
and much more difficult to construct it.

Below we present a radically
different method for verification of FPs 
based on a construction 
of state diagrams for FPs, 
and illustrate it by a proof of \re{4.21} 
on the base of this method.
This proof can be generated automatically,
that is an evidence of advantages of the
method for verification of FPs 
based on state diagrams.

\section{State diagrams of functional programs}

\subsection{Concepts and notations 
related to terms}

The following notations and
 concepts will be used below.
\bi
\i ${\cal E}$ is a set 
of all terms. 
\i ${\cal E}_0$ is a set of all terms
 not containing functional variables.
 \i ${\cal E}_{conc}$ is
a set of terms $e\in {\cal E}_0$, 
such that each FS occurred in $e$
is  $conc$.
\i If $\Sigma$ is a FP, then
${\cal E}_{\Sigma}$ is
a set of terms, each of which 
is either a variable or
has the form
 $\varphi(u_1,\ldots, u_n)$,
where 
$\varphi \in \Phi_\Sigma$ and $u_1,\ldots, u_n \in {\cal E}_{conc}$.
\i If $e\in {\cal E}$, then
$X_e$ and $\Phi_e$ 
are sets of
all data variables and functional
variables respectively occurred in  $e$.
\i If
   $e\in {\cal E}$,
   $x_1,\ldots, x_n$ is a list of the different variables, and
   $e_1,\ldots,e_n$ are terms such that
   $\forall\,i=1,\ldots, n$
      $type(e_i)=type(x_i)$,
   then the notation
   \be{cv33xzcvxzc}e(e_1/x_1,\ldots, e_n/x_n)\ee
   denotes a term derived from $e$
  by replacement 
   $\forall\,i\in \{1,\ldots,n\}$
   all occurrences of $x_i$ in $e$ 
   on the term $e_i$.
\i If $e$ and $e'$ are terms, 
then for each term $e''$, such that
$type(e'')=type(e')$,
the notation $e(e''/e')$ denotes
a term derived from $e$
by a replacement of all occurrences of
$e'$ in  $e$
   on the term $e''$.
\i An {\bf assignment}
is a notation of the form \be{dfdsafdsafsa}u:=e\ee where 
$u\in {\cal E}_{conc},\;\;  
 e\in {\cal E}_{\Sigma},\;\; type(u)=type(e).$ 
\i If $X\subseteq {\cal X}$, 
then an {\bf evaluation} of variables
occurred in 
$X$ is a function $\xi$, which maps
each variable $x \in X$ to a value
 $x^\xi\in {\cal D}_{type(x)}$.
 The set of all evaluations 
 of variables
occurred in 
 $X$ will be denoted by 
 $X^\bullet$.

\i 
For each  $e\in {\cal E}_0$, 
each $X\supseteq X_e$
and each $\xi\in X^\bullet$
the notation $e^\xi$
denotes an object called
{a \bf value} of $e$ on $\xi$
and defined by a standard way
(i.e. if $e\in {\cal C}$, then $e^\xi$ is equal
to the value of the constant $e$, 
if $e\in {\cal X}$, then $e^\xi$ is equal to the value of the evaluation
$\xi$ on the variable $e$,
and if
$e=f(e_1,\ldots, e_n)$,
then $e^\xi = f(e_1^\xi,\ldots, e_n^\xi)$).

\i We shall consider terms $e_1,e_2\in {\cal E}_0$ as equal iff for each 
$\xi\in (X_{e_1}\cup X_{e_2})^\bullet$ the equality
$e_1^\xi=e_2^\xi$ holds. 
We understand this equality 
in the following sense:
values
$e_1^\xi$ and $e_2^\xi$ 
either both undefined,
or both defined and coincide.

\i A term
$e\in {\cal E}_0$ is called a 
{\bf formula}, if all variables from
$X_e$ are of the type ${\bf C}$, and  
$\forall\,\xi\in  X_e^\bullet\quad
e^\xi\in \{0,1\}$.
The symbol ${\cal B}$ denotes the set of 
all formulas.
The symbols $\top$ and $\bot$ 
denote formulas taking the values
1 and 0 respectively
on each evaluation of their variables.
\ei

\subsection{A concept of a state of a FP}

Let $\Sigma$ be a FP.

A {\bf state} of $\Sigma$ 
is a notation $s$ of the form 
\be{fdgdsfghdsfgssdfgdf}b.u\,(\theta_1,\ldots,\theta_m)\ee
components of which are the following:
\bi
\i $b$ is a formula from ${\cal B}$,
called a {\bf condition}  of $s$,
\i $u$ is a term from ${\cal E}_{conc}$, called a {\bf value} of
 $s$,  and
\i $\theta_1,\ldots,\theta_m$ are
assignments.
\ei

We shall use the following notations.

\bi
\i $S_\Sigma$ is the set of all states of $\Sigma$.
\i If a state $s\in S_\Sigma$ 
is of the form
\re{fdgdsfghdsfgssdfgdf}, then
we shall denote by
$b_s$, $u_s $, $\Theta_s$ and $type(s)$
a formula ${b}$, a term $u$,
a sequence of assignments (which 
can be empty)
in \re{fdgdsfghdsfgssdfgdf}, and 
a type $type(u)$,
respectively.

If $b_{s}=\top$, then
the formula ${b}$ in
 \re{fdgdsfghdsfgssdfgdf}
will be omitted.

\i If $s\in S_\Sigma$, then \bi \i
 $X_s$ is a set of all data variables
occurred in $s$,
\i each variable from $X_s$, occurred in
the left side of an assignment from
$\Theta_s$, is called an 
{\bf internal variable} of $s$, 
 all other variables from $X_s$ are
 called
 {\bf input variables} of $s$,
 \i $s^\bullet$ is a set
of all $\xi\in X_s^\bullet$,
such that 
$b_{s}^\xi=1$, and $\forall\,(u_i:=e_i)\in \Theta_s$
\bi
\i if $e_i\in {\cal E}_{conc}$, then 
$u_i^\xi=e_i^\xi$, and
\i if $e_i=\varphi(v_1,\ldots, v_n)$, then 
$$u_i^\xi=f_{\varphi}(v^\xi_1,\ldots, v^\xi_n),$$
where $f_{\varphi}$ is a corresponding component of a LFP of $\Sigma$.
\ei\ei
\ei

A state $s\in S_\Sigma$ is said to be 
{\bf terminal}, 
if  $\Theta_s$ does not contain 
functional variables.

Given a pair of states  
$s_1,s_2\in S_{\Sigma}$.
We denote by the notation
$s_1\subseteq  s_2$ the following
statement: sets of input variables 
 $s_1$ and $s_2$ are equal, and
$$
\forall\,\xi_1\in s_1^\bullet\;\exists \,\xi_2
\in s_2^\bullet: u_{s_1}^{\xi_1}=u_{s_2}^{\xi_2}.$$

Along with the states of  FPs, 
we shall consider
also {\bf pseudo-states}, which
differ from states only that
their assignments have the form 
$u: = e$, where $u\in {\cal E}_{conc}$,
$e\in {\cal E}$.
For each pseudo-state $s$ 
the notations $b_s$, $u_s$
and $\Theta_s$
have the same meaning as for states.

\subsection{Unfoldinig of states}
\label{sadfsadfasd3334}

Let $\Sigma$ be a FP,
$s\in S_\Sigma$ be a state, 
$\theta\in \Theta_s$ be an assignment 
of the form
$$u:=\varphi(v_1,\ldots, v_n)$$
and an equation in $\Sigma$
that corresponds to $\varphi$
has the form 
$\varphi(x_1,\ldots, x_n)=e_\varphi.$

Denote by 
$s^\theta$ a set, 
called an {\bf unfolding} of the 
state $s$ with respect to $\theta$,
and defined by 
the procedure of its construction,
which consists of the steps listed below.

\begin{description}
\i[Step 1.]$\;$\\
$s^\theta$ is assumed to be a singleton,
which consists of a pseudo-state, 
derived from $s$ by a replacement of
$\theta$ on the assignment
$$u: =e_\varphi(v_1/x_1,\ldots, v_n/x_n).$$

\i
[Step 2.]$\;$\\
(This step can be performed
several times until there is the possibility
to perform it.)\\
  If all the elements of the set 
  $s^\theta$ are states from 
$S_\Sigma$, 
then the performance of this
step ends, otherwise
$s^\theta$ is modified
in the following way.

We choose an arbitrary element
$s'\in s^\theta$, which is not
a state of $S_\Sigma$, and denote by
$\theta'$ the first of the assignments,
occurred in $\Theta_{s'}$, 
 which has the form $u:=e$,
where $e\not\in {\cal E}_\Sigma$. 
Consider all possible variants 
of the form of the term $e$,
and for each of these variants, 
we present a rule of a modification 
of the set $s^\theta$,
according to this variant.
Below, the phrase ``a new variable''
means ``a variable that has no occurrences in the pseudo-state
under consideration''.

\bi
\i $e\in {\cal C}$, in this case
\bi
\i if $u=e$, then
remove $\theta'$ from $s'$,
\i if $u\in {\cal X}$, then replace
all occurrences of $u$ in $s'$  on 
$e$, and remove $\theta'$ from $s'$,
\i otherwise
remove $s'$ from $s^\theta$.
\ei
\i $e=e'_h$, in this case replace $\theta'$ 
on the assignment
\bi
\i  $u := e_1$, 
if $e'$ has the form  $e_1e_2$,
\i  $ux :=e'$, where $x$ is  a new variable,
otherwise.
\ei

\i $e=e'_t$, in this case replace  $\theta'$
on the assignment
\bi
\i $u := e_2$, 
if $e'$ has the form $e_1e_2$,
\i  $xu :=e'$, where $x$ is
a new variable, otherwise.
\ei

\i $e=e_1e_2$, in this case  \bi\i 
if  $u=u_1u_2$, then 
replace  $\theta'$ on
   a couple of assignments
    $u_1 :=e_1$, $u_2 :=e_2$,   
   \i if $u\in {\cal X}$, then 
   replace all occurrences of  $u$ in 
   $s'$ on the term 
    $xy$ (where $x$ and $y$ are new variables), and  
   $\theta'$ on
the couple of assignments
 $x :=e_1$, $y :=e_2$,   
 \i otherwise remove $s'$ from
  $s^\theta$.\ei

\i $e=(e_1=\varepsilon)$, in this case 
   \bi
   \i add to $s^\theta$ a copy of the state 
   $s'$ (denote it  by $s''$),
   \i replace
     \bi\i $\theta'$ in $s'$  on the couple   $u:=1$,
     $\varepsilon: = e_1$,
     and \i $\theta'$ in $s''$   on the couple $u:=0$,
     $xy := e_1$, where $x$ and  $y$ are new variables.\ei
   \ei   

\i  $e=(e_1=e_2)$, $e=(e_1\leq e_2)$,
$e=(e_1\wedge e_2)$ etc.,
in this case  \bi\i replace $\theta'$
   on the couple
      $x_1 := e_1$, $x_2 := e_2$, 
   where $x_1$, $x_2$ 
   are new variables, 
and \i add to $b_{s'}$
the conjunctive member 
$u=e'$, 
where $e'$ is derived from 
$e$ by a replacement of $e_i$ 
on $x_i\;(i=1,2)$.\ei
 
\i $e=(e_1? e_2:e_3)$,  in this case add
   to $s^\theta$ 
   a copy of  $s'$ (denote it by $s''$), and 
   replace all occurrences 
   \bi\i $\theta'$ in $s'$ on the couple 
     $1: =e_1$, $u :=e_2$, \i
     $\theta'$ in $s''$ on the couple
       $0: =e_1$, $u :=e_3$.\ei

\i $e=\varphi(e_1,\ldots, e_k)$, 
$\exists\,i:
e_i\not\in {\cal E}_{conc}$,
in this case, replace
$e_i$ in $\theta'$ on the new variable 
$x$,
and add $x :=e_i$ before $\theta'$.
\ei

 \i [Step 3.]  $\;$\\
For each $s'\in s^\theta$
\bi
\i if 
$\Theta_{s'}$ has a pair
of the form
$u:=x$, $v:=x$, where 
$x\in {\cal X}$, and $u,v$ are of the form
$u_1\ldots u_n$, $v_1\ldots v_m$ respectively, then there is executed
an algorithm which consists of
the following steps: \\
(as a result of 
each of the these steps it is changed
a form of these assignments,
but we will denote the changed assignments by the same notation
as  original assignments):

\bi
\i if $n<m$, then in the case $u_n\in {\cal X}$  each occurrence of the variable $u_n$ in $s'$  is replaced on the term
$v_n\ldots v_m$, and in the case $u_n=\varepsilon$ we remove $s'$ from $s^\theta$,
\i analogously in the case $m<n$, 
\i $\forall\,i=1,\ldots, n$:
\bi
\i if $u_i\in {\cal X}$, then replace all occurrences $u_i$ in $s'$ on $v_i$, and if $u_i\not\in {\cal X}$, but
$v_i\in {\cal X}$, then replace all occurrences $v_i$ in $s'$ on $u_i$,
\i if $u_i\neq v_i$,  then remove   $s'$ from $s^\theta$,
\ei
\i delete one of the considered assignments,
\ei
\i if $b_{s'}=\{b',x=u\}$, where $x\in {\cal X}$, $u\in {\cal X}\cup {\cal C}$, then 
$b_{s'}$ is replaced on $b'$,  and all occurrences $x$ in $s'$  are replaced on $u$,
\i $b_{s'}$ is simplified by 
\bi \i a replacement
of subterms without variables 
to corresponding constants, and \i
simplifying 
 transformations related to 
 boolean identities and 
 properties of 
equality and linear
order relations, \ei
\i if $b_{s'} = \bot$, 
then $s'$ is removed from $s^\theta$.
\ei

\end{description}

{\bf Theorem 1}.

The above procedure for constructing of 
the set $S^\theta$ is always terminated.
$\blackbox $\\

A state $s\in S_\Sigma$ is 
{\bf inconsistent},
if it is not terminal, and
$\exists \,\theta\in \Theta_s$: either
 $s^\theta=\emptyset$, or
all  states in $s^\theta$ are inconsistent.

\subsection{Substitution of states in terms}

Let $\Sigma$ be a FP, 
$e$ be a term, $x_1,\ldots, x_n$
be a list of 
different variables from ${\cal X}$, 
and $s_1,\ldots, s_n$ be a list of states from
$S_\Sigma$, such that
  $\forall\,i=1,\ldots, n\;\;
type(s_i)=type(x_i)$.
The notation
\be{cv33433xzcvxzc}e(s_1/x_1,\ldots, s_n/x_n)\ee
denotes a state $s_e\in S_\Sigma$, defined
by induction on the structure of  $e$:
\bi
\i if $e=x_i\in \{x_1,\ldots, x_n\}$, then $s_e\eam s_i$,
\i if $e\in {\cal X}\setminus\{x_1,\ldots, x_n\}$
or $e\in {\cal C}$, then
$s_e\eam e\,(\,)$,
\i if $e=g(e_1,\ldots,e_k)$, where $g\in {\cal F}\cup \Phi$,
and the states $s_{e_1},\ldots, s_{e_k}$
of the form \re{cv33433xzcvxzc}, 
which are corresponded to 
terms $e_1,\ldots,e_k$, are already defined,
then $s_{e}$ is defined as follows:
\bi
\i internal variables of the states $s_{e_i}$ 
are replaced on new variables 
by a standard way, so that
all the internal variables 
of these states will be
different,
let $b_i.u_i(\Theta_i)\;(i=1,\ldots, k),$
be the
resulting
states,
\i $s_e$ is a result of an application of 
actions 2 and 3 from section
\re{sadfsadfasd3334}
to the state
$$\{b_1,\ldots, b_k\}.g(u_1,\ldots, u_k)\,
(\Theta_1,\ldots, \Theta_k).$$
\ei
\ei

Term  \re{cv33433xzcvxzc}
will be denoted by the notation
$e(s_1,\ldots, s_n)$, in that case,
when the list of the variables 
$x_1,\ldots, x_n$ is clear from the context.

\subsection{A concept of a state diagram 
of a FP}

Let $\Sigma$ be a FP, and 
left side of first equation in $\Sigma$
has the form $\varphi(x_1, \ldots, x_n)$.
 
A {\bf state diagram (SD)} of the FP
$\Sigma$ is a graph $G$
with distinguished node $n_0$
(called an {\bf initial node})
satisfying the following conditions.
\bi
\i Each node $n$ of the graph $G$
is labelled by a state 
$s_n \in S_\Sigma$,
and $s_{n_0} $ has the form
$$y\,(y:=\varphi(x_{1},\ldots, x_{n})),
\quad\mbox{where $y\not\in \{x_{1},\ldots, x_{n}\}$.}$$
\i For each node $n$ of the graph $G$
one of the following
statements holds.
\bn
\i There is no an edge outgoing from $n$,
and $s_n$ is terminal.
\i There are two edges outgoing from $n$,
 and states $s',s''$ corresponded to ends
of these edges
have the following property:
$\exists\,x\in X_{s_n}: type(x)={\bf S}$,
there are no assignments of the form $u:=x$
in $\Theta_{s_n}$, and $s',s''$ 
 are obtained from ${s_n}$ by
\bi \i a replacement of all 
occurrences of $x$ 
by the constant $\varepsilon$ and by 
the term $yz$ respectively
(where $y$ and $z$ are 
variables not occurred in $X_{s_n}$), and
\i if $x$ is not occurred 
in the left side of any assignment 
from  $\Theta_{s_n}$, 
then -- by adding
assignments 
$\varepsilon: = x$ and $yz: = x$
to $\Theta_{s'}$ and $\Theta_{s''}$
respectively. \ei
\i $\exists\,\theta\in \Theta_{s_n}$:
a set of states corresponding to
 ends of  edges outgoing
from $n$, is equal to the set of all
consistent states 
 from $s_n^\theta$.
\i $u_{s_n}$ has the form $u_1u_2$, 
and there is one edge outgoing from 
$n$ labeled by $tail$,
and the end $n'$ of this edge 
satisfies the condition:
$tail(s_n)\subseteq s_{n'}.$
\i There is an edge outgoing from 
$n$ labelled by $<$, the end $n'$ of which
satisfies the condition:
\bi\i $\exists\, n_1,n_2$: 
$G$ contains an edge from $n_1$ to $n_2$
labelled by $tail$, and
\i $\exists\,e\in {\cal E}_\Sigma,\;\exists \,x\in X_e:$
$$s_n\subseteq e(tail(s_1)/x),\quad e(s_2/x)\subseteq s_{n'}.$$
\ei
\en
\ei

We describe an informal sense 
of the concept of a SD.
Each state $s$ can be considered as
a description of a process of
a calculation of the value $u_{s}$
on concrete values 
of input variables of this state
(by an execution of 
assignments from $\Theta_{s}$,
checking the condition $b_{s}$ 
and a calculation of the value of the term
$u_{s}$ on the calculated values ??of the variables occurred in this term). 
If all edges outgoing from the state
$n$ are unlabeled, then
ends of these edges correspond 
to possible
options for calculating the value of $u_{s_n}$
(by detailization of a structure 
of a value of some variable 
from $X_{s_n}$,
or by an equivalent transformation
of any assignment from $\Theta_{s_n}$).
If there is an edge from $n$ to $n'$
labeled by $tail$, then this edge 
expresses a reduction of the problem of calculating of the tail of 
the value 
 $u_{s_n}$ to the problem
 of calculating the value of $u_{s_{n'}}$.
If there is an edge from $n$ to $n'$
labeled by $<$, then this edge 
expresses a reduction of the problem of calculating the value
$u_{s_n}$ to the problem 
of calculating the value $u_{s_{n '}}$
on arguments on the smaller size.

We say that FP $\Sigma$
{\bf has a finite SD}, if there
is a SD of $\Sigma$ with finite set of nodes. \\

{\bf Theorem 2}.

Let
 $\Sigma_1$ and $\Sigma_2$
have finite SDs,
$\Phi_{\Sigma_1}\cap \Phi_{\Sigma_2} = \emptyset$, and
 left sides of  first equations 
in $\Sigma_1$ and $\Sigma_2$
have the form
$\varphi_1(x_1,\ldots, x_n)$ and 
$\varphi_2(y_1,\ldots, y_m)$ 
respectively, where 
$type(\Sigma_1)=type(y_1)$.

Then
FP $\Sigma$ such that
\bi\i its 
first equation has the form
$$\by\varphi(x_1,\ldots, x_n,y_2,\ldots, y_m)=\\=
\varphi_2(\varphi_1(x_1,\ldots, x_n), y_2,\ldots, y_m)\ey$$
\i and a set of other
equations is 
$\Sigma_1 \cup \Sigma_2$ 
\ei
has a finite SD. $\blackbox $\\
 
We do not give a description of the algorithm for the construction of a finite SD for 
$\Sigma$
 due to limitations on the size of the article.
  We note only that the
 SD is a 
union of a SD for 
$\Sigma_1$,
 a SD for $\Sigma_2 $, and a SD, which is a Cartesian product of two
 previous  SDs. \\
 
{\bf Theorem 3}.

Let FP $\Sigma$ has a finite
SD, where
values of
states corresponding to
terminal nodes of this SD, which
are reachable from an
initial state,
are equal to $1$.
Then $f_\Sigma$
has value 1 on all its arguments.
$\blackbox $\\

The above theorems are
theoretical foundation 
of new method of verification of FPs.
This method consists in a 
constructing finite SDs
\bi\i for a FP $\Sigma_1$ under verification, 
and \i for a FP $\Sigma_2$ which represents
some property of $\Sigma_1$.\ei
If there are finite SDs for $\Sigma_1$ and 
$\Sigma_2$, then, 
according to Theorem 2, there is a 
finite SD for
a superposition of $\Sigma_1$ and $\Sigma_2$.
If this SD has the property
indicated in Theorem 3, 
then the superposition
of functions corresponding 
to $\Sigma_1$ and $\Sigma_2$,
has the value 1 on all its arguments.

In the next section we present
an example of this method.

For a constructing of SDs it is used
a method of justification of 
statements of the form 
$s_1 \subseteq s_2$, 
which we did not set out here
due to limitations on the size of the article.
We only note that this method uses the concept of an unification of terms.

We shall use the following convention
for graphical presentation of SDs:
if a state $s$ associated with a node of a SD
has the form
$b.u
(\theta_{1},\ldots,\theta_{n})$,
then this node is designated by an oval,
 over which it is drawn a notation $b.u$
 (or $u$, if $b = \top$),
 and components of the list $\Theta_{s}$
 are depicted inside the oval.
 An identifier of the node 
 can be depicted from the left of the oval.

  which are listed in
Components list $\Theta_{s} $, and
 the left of which might be painted symbol
It is
ID of this node.

\section {An example of verification
of a FP by constructing of a state diagram}

In this section
we illustrate the verification method
outlined above by an
example of verification of FP of 
sorting, in this case $\Sigma_1 =$
\re{fdsgdsgdsr} and $\Sigma_2 =$
\re{sdfgfdsgsdfgdsfgsrrr}.

We shall use the following convention:
if nodes $n_1$ and $n_2$ of a SD
are such that $n_2$ can be derived from 
$n_1$ 
by a performing of actions 2 and 3 from 
the definition of a SD, 
then we draw an unlabeled
edge from $n_1$ to $n_2$
(i.e. unlabeled edges in a new 
understanding of a SD correspond to
paths consisting of unlabeled edges in
original understanding of a SD).

\subsection {A state diagram for the
 FP of sorting}

In this section we describe the process of building of a
SD for FP \re{fdsgdsgdsr}.
Terms of the form ${\bf insert}(a,y)$
we denote by $a\to y$.

An initial node of the SD
for FP
\re{fdsgdsgdsr}
(this node will be denoted by the symbol
$A$)
has the form \\$$
\put(0,20){\makebox(0,0){$
y$}}
\put(0,0){\oval(70,20)}
\put(0,0){\oval(74,24)}
\put(0,0){\makebox(0,0){$
y:={\bf sort}
(x)$}}
\put(-42,10){\makebox(0,0){$
A$}}$$

Two unlabeled 
edges can be drawn 
(corresponding to 
replacement of $x$ by $\varepsilon$
and by $ab$, and to an unfolding 
of one assignment)
from this node to the nodes
$$
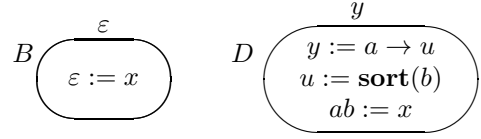
\put(-50,0){\oval(50,30)}
\put(-50,0){\oval(50,30)}
\put(-50,0){\makebox(0,0){$\bcy
\varepsilon:=x
\ey$}}
\put(-50,20){\makebox(0,0){$\bcy
\varepsilon
\ey$}}
\put(-80,10){\makebox(0,0){$
B$}}
\put(50,0){\oval(80,40)}
\put(50,0){\makebox(0,0){$\bcy
y:=a\to  u\\
u := {\bf sort}(b)\\
ab:=x
\ey$}}
\put(45,26){\makebox(0,0){$y
$}}
\put(2,10){\makebox(0,0){$
D$}}$$

Also it is possible to draw two unlabeled 
edges (corresponding to replacement 
of the variable
 $u$ by constant $\varepsilon$
and by the term $cd$)
from $D$ to nodes with labels
\be{dfsadfdsafs}y\,(y:=a\to cd,cd:={\bf sort}(b),ab:=x),
\ee
and
\be{d3fsadfdsafs}y\,(y:=a\varepsilon,\varepsilon:={\bf sort}(b),ab:=x).
\ee
Also it is possible to draw two edges
from the node labeled by \re{dfsadfdsafs}
to nodes labeled by 
$$\by C:\quad \{a\leq c\}.acd\,(cd:={\bf sort}(b),ab:=x),\\
G: \quad \{c<a\}.cz\,(z:=a\to d,cd:={\bf sort}(b),ab:=x)
\ey$$
(by an unfolding of the first assignment).

It is possible to draw an edge labeled by 
$tail$ from $C$ to the initial node
(the existence of such an edge is seen directly).

It is possible to draw two edges
from the node labeled by \re{d3fsadfdsafs}
(replacing $b$ to $\varepsilon $ and
to $pq$)  to nodes, 
one of which is terminal and has the form
$$E:\quad a\varepsilon\;(a\varepsilon:=x),$$
and the second node is inconsistent (that
can be determined 
by additional unfoldings,
which we do not present here).

It is possible to draw two unlabeled edges
from $G$
(corresponding to the
replacement of $b$ to $\varepsilon $ and
to $pq$)  to nodes, one of which is 
inconsistent, and the second node
is labeled by
\be{dsafasdfdsaf}
\{c<a\}.\,cz\,\left(\by z:=a\to d,\\ cd:=p\to w, \\w:={\bf sort}(q),\\
apq:=x\ey\right).\ee
It is possible to draw two unlabeled edges
from \re{dsafasdfdsaf} 
(corresponding to the replacement 
of $w$ to $\varepsilon$
and $ij$):
\bi \i from the end of the first of these edges it can be drawn several unlabeled edges,
but among ends of all these edges
there is a unique 
consistent node labeled by
$$\{c<a\}.\,cz\,\left(\by z:=a\to \varepsilon,\\c:=p,\\d:=\varepsilon,\\ap\varepsilon:=x\ey\right),$$
and there is a unique unlabeled edge 
from this node
to a terminal node 
$$H:\quad \{c<a\}.\,ca\varepsilon\,
(ac\varepsilon:=x),$$
\i the end of second edge has a label
\be{dfs44434dfsdfsa}\{c<a\}.\,cz\,\left(\by z:=a\to d,\\
cd:=p\to ij, \\ij:={\bf sort}(q),\\apq:=x\ey\right).\ee
\ei

It can be drawn a couple of edges
from \re{dfs44434dfsdfsa},
the ends of which have labels
$$F:\quad
\{c<a,c\leq i\}.\,cz\,\left(\by z:=a\to ij,\\
ij:={\bf sort}(q),\\acq:=x\ey\right),$$
$$I:\quad
\{c<a,c<p\}.\,cz\,\left(\by z:=a\to d,\\
d:=p\to j, \\cj:={\bf sort}(q),\\apq:=x\ey\right).$$

It can be drawn an edge labeled by $tail$ 
from $F$ to the initial node (the existence 
of such an edge is seen directly).

A pair of nodes $(D, I)$ is related to 
the pair of nodes
$(A, G)$ by the following relations:
  \be{adsfasdfasdfsa}tail(I)=e(tail(G)/h),\quad D=e(A/h)\ee 
  where $e=a\to h$. 
  In other words, labels of nodes 
$I, D$ can be obtained from labels of nodes 
$G, A$ by adding 
an assignment to the top.
This fact can be used to justify
an existence of an edge from $G$ to $A$
with label $tail$.
We do not set out the detailed justification 
of an existence of such an edge, 
we describe only 
a scheme of such a justification.
Let $\rho(x)$ be a partial function 
with the following property: 
if $\rho$ is defined 
on a value $\alpha $ of the variable $x$, 
then it maps
$\alpha$ to a string 
$\beta$, which has the property
$$u_{tail(G)}^{x\mapsto \alpha}=u_A^{x\mapsto\beta}.$$
The formula 
\re{adsfasdfasdfsa} directly implies
the following
property of the function $\rho$:
\be{dfasdfdsafadsa}x\neq \varepsilon\;\Rightarrow\;
\rho(x)\sqsupseteq x_h\rho(x_t)\ee
where the inequality $\sqsupseteq$ is
understood as an order relation on the
set of partial functions: if for some
value of the variable
$x$ the right side of \re{dfasdfdsafadsa}
is defined, then the left side also is
defined for this value of $x$, and
and values ??of both parts are the same.

A property of 
totality of the function $\rho$ is justified
by the inequality 
\re{dfasdfdsafadsa} and by an analysis
of a fragment of 
SD for \re{fdsgdsgdsr} which is already built.
Note that this justification can
be generated automatically.
A proof of correctness of this justification
is based on the 
concept of  unification of state pairs, 
it has a large volume, and we omit it.

The constructed SD for FP
\re{fdsgdsgdsr} is shown in Fig. 1,
it can be simplified to the SD in Fig. 2.

\subsection{A state diagram for 
the FP of cheking of string ordering}

A fragment of a SD for FP $\Sigma_2$ 
(see \re{sdfgfdsgsdfgdsfgsrrr})
(consisting of nodes reachable from the initial state) has the form 

$$\by\mbox{ $\;$ }
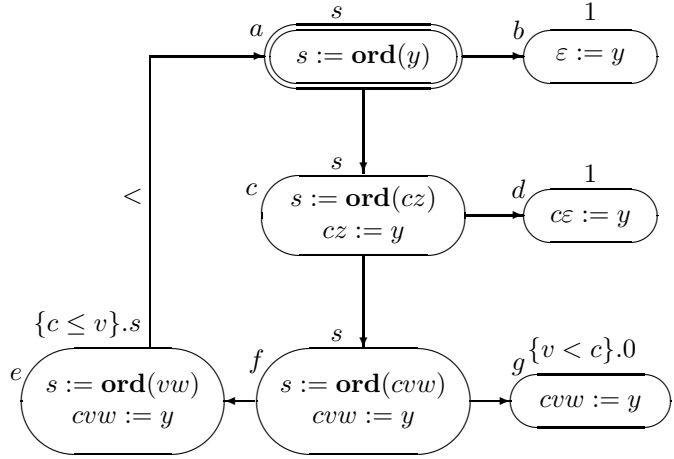
\begin{picture}(0,170)

\put(-40,160){\makebox(0,0){$a$}}
\put(58,160){\makebox(0,0){$b$}}
\put(-42,100){\makebox(0,0){$c$}}
\put(58,100){\makebox(0,0){$d$}}
\put(-130,30){\makebox(0,0){$e$}}
\put(-40,35){\makebox(0,0){$f$}}
\put(58,33){\makebox(0,0){$g$}}

\put(-10,167){\makebox(0,0){$\bcy 
s
\ey$}}

\put(0,150){\oval(74,24)}
\put(0,150){\oval(70,20)}
\put(0,150){\makebox(0,0){$\bcy 
s:={\bf ord}(y)
\ey$}}

\put(-10,110){\makebox(0,0){$\bcy 
s
\ey$}}

\put(0,90){\oval(76,30)}
\put(0,90){\makebox(0,0){$\bcy 
s:={\bf ord}(cz)\\cz:=y
\ey$}}

\put(85,167){\makebox(0,0){$
1$}}

\put(85,150){\oval(50,20)}
\put(85,150){\makebox(0,0){$
\varepsilon:=y$}}

\put(85,105){\makebox(0,0){$\bcy
1
\ey$}}

\put(85,90){\oval(50,20)}
\put(85,90){\makebox(0,0){$\bcy
c\varepsilon:=y
\ey$}}

\put(-10,45){\makebox(0,0){$\bcy
s\ey$}}

\put(0,20){\oval(80,40)}
\put(0,20){\makebox(0,0){$\bcy
s:={\bf ord}(c  v  w)\\
cvw:=y
\ey$}}

\put(85,20){\oval(60,20)}
\put(85,20){\makebox(0,0){$\bcy
cvw:=y
\ey
$}}

\put(-90,20){\oval(76,40)}
\put(-90,20){\makebox(0,0){$\bcy
s:={\bf ord}(v  w)\\
cvw:=y

\ey$}}

\put(0,138){\vector(0,-1){32}}
\put(0,75){\vector(0,-1){35}}
\put(-41,20){\vector(-1,0){11}}
   \put(-105,54){\makebox(0,0)[t]{
   $\{ c\leq v\}.s$}}
\put(40,20){\vector(1,0){15}}
   \put(80,44){\makebox(0,0)[t]{
   $\{ v<c\}.0$}}
\put(38,90){\vector(1,0){22}}

\put(37,150){\vector(1,0){23}}

\put(-80,40){\line(0,1){110}}
\put(-80,150){\vector(1,0){43}}

   \put(-88,100){\makebox(0,0)[t]{
   $<$}}

\end{picture}\\
\;\\
\ey\\\mbox{ }
$$

\subsection{A state diagram 
for a superposition of the sorting FP
and the FP of ordering checking}

There is an algorithm based on Theorem 3,
which can be applied to  SDs for the FPs
\re{fdsgdsgdsr} and \re{sdfgfdsgsdfgdsfgsrrr}, 
which results the SD shown in Fig. 3.
This SD has two terminal nodes, and
labels of both these nodes have a value of 1.
According to Theorem 3,
this implies that
the function ${\bf ord} \circ {\bf sort}$
has the value 1 on all its arguments.

In conclusion we note,
that despite on the complexity 
of all of the above
transformations and reasonings, 
all of them can be generated automatically.
An attempt to justify an existence 
of edges with labels $tail$ 
and $<$ can be executed
automatically for each pair
of nodes arising in the process 
of building of the SD.
It can be seen from this example 
that the process of a construction 
of a SD is terminated fast enough.

\section{Conclusion}

We have proposed the concept of a state diagram (SD) for functional programs (FPs) and a verification method based on the concept of 
a SD.
One of the problems for further research related to the concept of a SD 
has the following form: 
find a sufficient condition $\varphi$
(as stronger as possible) 
on a FP $\Sigma$ such that
if $\Sigma$ meets $\varphi$ then 
$\Sigma$  has a finite SD.

\newpage
Fig. 1:

$$\by
\begin{picture}(0,265)

\put(0,240){\oval(70,20)}
\put(0,240){\oval(74,24)}
\put(0,240){\makebox(0,0){$
y:={\bf sort}
(x)$}}

\put(120,260){\makebox(0,0){$\bcy
\varepsilon
\ey$}}

\put(120,240){\oval(50,30)}
\put(120,240){\makebox(0,0){$\bcy
\varepsilon:=x
\ey$}}

\put(0,180){\oval(80,40)}
\put(0,180){\makebox(0,0){$\bcy
y:=a\to  u\\
u := {\bf sort}(b)\\
ab:=x
\ey$}}

\put(-5,206){\makebox(0,0){$y
$}}

\put(-5,258){\makebox(0,0){$y
$}}

\put(120,200){\makebox(0,0){$a\varepsilon
$}}

\put(120,180){\oval(50,30)}
\put(120,180){\makebox(0,0){$\bcy 
a\varepsilon:=x\ey
$}}

\put(0,115){\oval(80,54)}
\put(0,115){\makebox(0,0){$\bcy
z := a\to d\\
cd:={\bf sort}(b)\\
ab:=x
\ey$}}

\put(82,146){\makebox(0,0){$tail$}}
\put(70,68){\makebox(0,0){$tail$}}
\put(-60,220){\makebox(0,0){$tail$}}
\put(-74,155){\makebox(0,0){$tail$}}
\put(-110,240){\makebox(0,0){$\{ a\leq c \}.acd$}}
\put(-30,148){\makebox(0,0){$\{ c<a \}.cz$}}

\put(0,45){\oval(80,60)}
\put(0,45){\makebox(0,0){$\bcy
z:=a\to  d\\
d := p\to j\\
cj:={\bf sort}(q)\\
apq:=x
\ey$}}

\put(-120,130){\oval(70,60)}
\put(-120,130){\makebox(0,0){$\bcy
z:=a\to  ij\\
ij:={\bf sort}(q)\\
acq:=x
\ey$}}

\put(-125,210){\oval(70,44)}
\put(-125,210){\makebox(0,0){$\bcy
cd := {\bf sort}(b)\\
ab:=x
\ey$}}

\put(135,134){\makebox(0,0){$\{ c<a \}.ca\varepsilon$}}

\put(120,110){\oval(50,30)}
\put(120,110){\makebox(0,0){$\bcy
ac\varepsilon:=x
\ey$}}

\put(0,228){\vector(0,-1){28}}
\put(0,160){\vector(0,-1){18}}
\put(37,240){\vector(1,0){58}}
\put(40,180){\vector(1,0){55}}
\put(40,110){\vector(1,0){55}}
\put(0,88){\vector(0,-1){13}}
\put(-40,180){\vector(-2,1){50}}
\put(-90,212){\vector(2,1){53}}

\put(-40,110){\vector(-2,1){45}}
\put(-85,142){\vector(2,1){50}}

\put(70,199){\vector(-1,1){35}}
\put(70,143){\line(0,1){56}}
\put(40,113){\line(1,1){30}}

\put(70,129){\vector(-1,1){36}}
\put(70,80){\line(0,1){49}}
\put(40,50){\line(1,1){30}}

\put(-30,257){\makebox(0,0){$A$}}
\put(95,260){\makebox(0,0){$B$}}
\put(-150,237){\makebox(0,0){$C$}}
\put(-40,200){\makebox(0,0){$D$}}
\put(95,200){\makebox(0,0){$E$}}
\put(-43,135){\makebox(0,0){$G$}}
\put(-155,157){\makebox(0,0){${F}$}}
\put(-45,60){\makebox(0,0){${I}$}}
\put(95,130){\makebox(0,0){${H}$}}

\put(-107,167){\makebox(0,0){$\{ c<a,c\leq i
\}.cz$}}

\put(-55,80){\makebox(0,0){$\{ c<a,c<p
\}.cz $}}

\end{picture}
\ey
$$

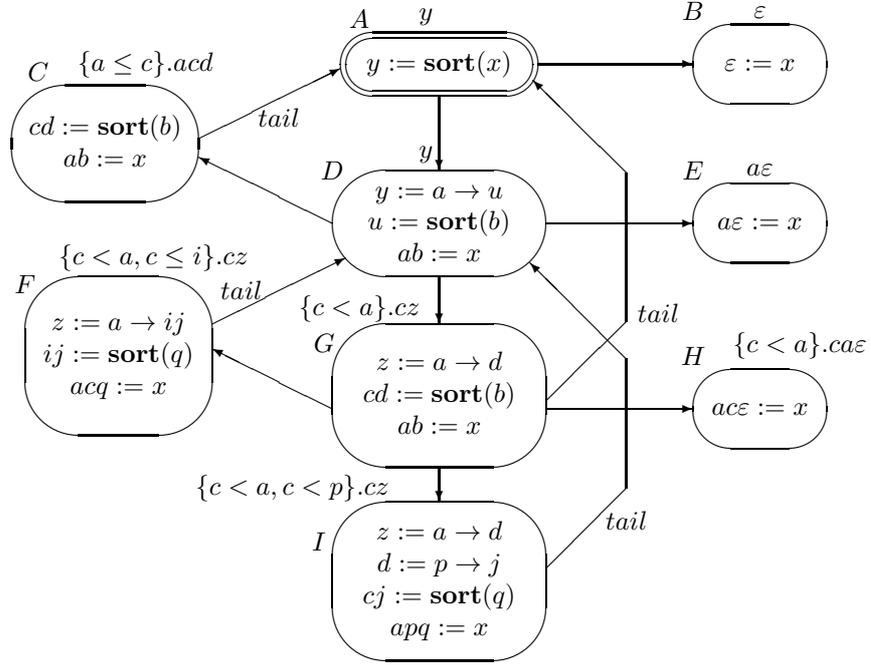
Fig. 2:\\

$$\by
\begin{picture}(0,165)

\put(0,140){\oval(70,20)}
\put(0,140){\oval(74,24)}
\put(0,140){\makebox(0,0){$
y:={\bf sort}
(x)$}}

\put(120,160){\makebox(0,0){$\bcy
\varepsilon
\ey$}}

\put(120,140){\oval(50,30)}
\put(120,140){\makebox(0,0){$\bcy
\varepsilon:=x
\ey$}}

\put(0,80){\oval(80,40)}
\put(0,80){\makebox(0,0){$\bcy
y:=a\to  u\\
u := {\bf sort}(b)\\
ab:=x
\ey$}}

\put(-5,106){\makebox(0,0){$y
$}}

\put(-5,158){\makebox(0,0){$y
$}}

\put(120,100){\makebox(0,0){$a\varepsilon
$}}

\put(120,80){\oval(50,30)}
\put(120,80){\makebox(0,0){$\bcy 
a\varepsilon:=x\ey
$}}

\put(0,20){\oval(80,44)}
\put(0,20){\makebox(0,0){$\bcy
z := a\to d\\
cd:={\bf sort}(b)\\
ab:=x
\ey$}}

\put(82,46){\makebox(0,0){$tail$}}
\put(-60,120){\makebox(0,0){$tail$}}
\put(-110,140){\makebox(0,0){$\{ a\leq c \}.acd$}}
\put(-30,48){\makebox(0,0){$\{ c<a \}.cz$}}

\put(-125,110){\oval(70,44)}
\put(-125,110){\makebox(0,0){$\bcy
cd := {\bf sort}(b)\\
ab:=x
\ey$}}

\put(0,128){\vector(0,-1){28}}
\put(0,60){\vector(0,-1){18}}
\put(37,140){\vector(1,0){58}}
\put(40,80){\vector(1,0){55}}
\put(-40,80){\vector(-2,1){50}}
\put(-90,112){\vector(2,1){53}}

\put(70,99){\vector(-1,1){35}}
\put(70,43){\line(0,1){56}}
\put(40,13){\line(1,1){30}}

\put(-30,157){\makebox(0,0){$A$}}
\put(95,160){\makebox(0,0){$B$}}
\put(-150,137){\makebox(0,0){$C$}}
\put(-40,100){\makebox(0,0){$D$}}
\put(95,100){\makebox(0,0){$E$}}
\put(-43,35){\makebox(0,0){$G$}}

\end{picture}
\ey
$$

\newpage
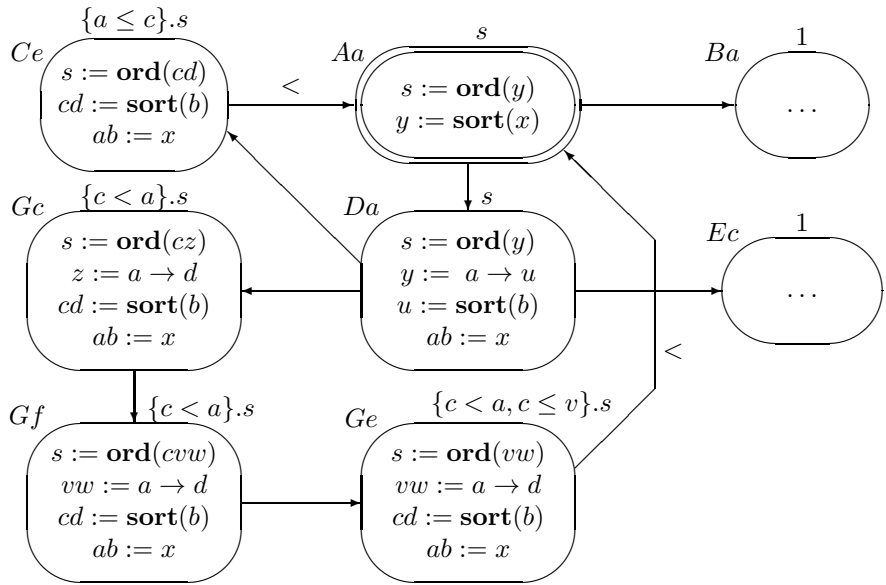
Fig. 3:\\

$$
\by
\begin{picture}(0,235)

\put(-45,210){\makebox(0,0){$Aa$}}
\put(95,210){\makebox(0,0){$Ba$}}
\put(-165,210){\makebox(0,0){$Ce$}}
\put(-165,152){\makebox(0,0){$Gc$}}
\put(-40,152){\makebox(0,0){$Da$}}
\put(95,142){\makebox(0,0){$Ec$}}
\put(-165,72){\makebox(0,0){$Gf$}}
\put(-40,72){\makebox(0,0){$Ge$}}

\put(125,215){\makebox(0,0){$\bcy
1
\ey$}}
\put(125,190){\oval(50,40)}
\put(125,190){\makebox(0,0){$\bcy
\ldots
\ey$}}

\put(125,145){\makebox(0,0){$\bcy
1
\ey$}}
\put(125,120){\oval(60,40)}
\put(125,120){\makebox(0,0){$\bcy
\ldots
\ey$}}

\put(40,120){\vector(1,0){55}}

\put(0,190){\oval(80,40)}
\put(0,190){\oval(84,44)}
\put(0,190){\makebox(0,0){$\bcy 
s:={\bf ord}(y)\\
y :={\bf sort}(x)
\ey$}}

\put(0,168){\vector(0,-1){18}}

\put(0,120){\oval(80,60)}
\put(0,120){\makebox(0,0){$\bcy 
s:={\bf ord}(y)\\
y := \;a\to u\\
u :={\bf sort}(b)\\
a  b :=x
\ey$}}

\put(-40,130){\vector(-1,1){50}}

\put(-40,120){\vector(-1,0){45}}
\put(-125,90){\vector(0,-1){20}}

\put(42,190){\vector(1,0){57}}

\put(-90,190){\vector(1,0){47}}

\put(-125,120){\oval(80,60)}
\put(-125,120){\makebox(0,0){$\bcy
s:={\bf ord}(c  z)\\
z := a\to  d\\
c  d  := {\bf sort}(b)\\
a  b :=x\ey$}}

\put(0,40){\oval(80,60)}
\put(0,40){\makebox(0,0){$\bcy 
s:={\bf ord}(v  w)\\
v  w := a\to  d\\
c  d  := {\bf sort}(b)\\
a  b :=x
\ey$}}

\put(-85,40){\vector(1,0){45}}

\put(-125,40){\oval(80,60)}
\put(-125,40){\makebox(0,0){$\bcy
s:={\bf ord}(c  v  w)\\
v  w := a\to  d\\
c  d  := {\bf sort}(b)\\
a  b :=x
\ey$}}

\put(-125,190){\oval(70,50)}
\put(-125,190){\makebox(0,0){$\bcy
s:={\bf ord}(c  d)\\
c  d  := {\bf sort}(b)\\
a  b :=x
\ey$}}

\put(-125,222){\makebox(0,0){$\{ a\leq c\}.s $}}
\put(20,77){\makebox(0,0){$\{ c<a, c\leq v\}.s $}}
\put(-100,75){\makebox(0,0){$\{ c<a\}.s $}}
\put(-125,155){\makebox(0,0){$\{ c<a\}.s $}}

\put(7,155){\makebox(0,0){$s $}}
\put(5,217){\makebox(0,0){$s $}}

\put(70,139){\vector(-1,1){34}}
\put(70,83){\line(0,1){56}}
\put(40,53){\line(1,1){30}}

\put(-68,200){\makebox(0,0)[t]{   $<$}}

\put(75,100){\makebox(0,0)[t]{   $<$}}

\end{picture}
\ey
$$

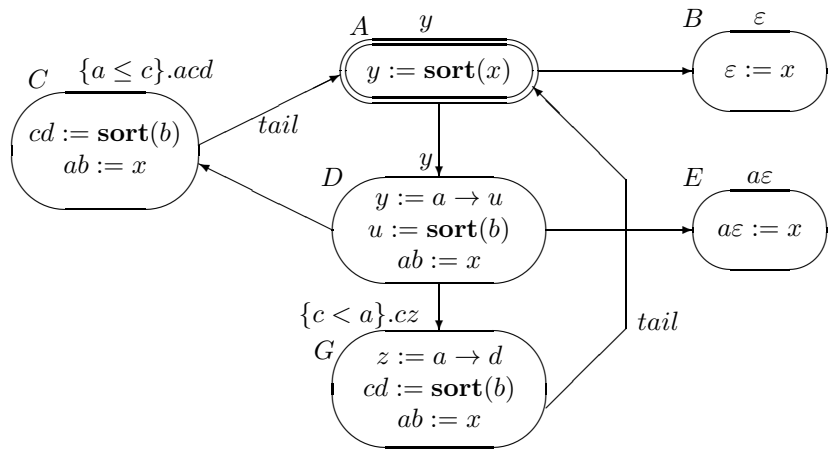
\end{document}